I.V. Manina, N.M. Peretolchina, N.S. Saprikina, A.M. Kozlov, I.N. Mikhaylova, A.Yu. Barishnikov

# EXPERIMENTAL RESEARCHES OF CUTANEOUS MELANOMA IMMUNOTHERAPY BY ANTITUMOR CELL-WHOLE GM-CSF-PRODUCING VACCINES

*N.N. Blokhin Russian Cancer Resrarch Center of RAMS, Moscow*

**\* Correspondence should be addressed to:**

Irina Manina,
N.N. Blokhin Russian Cancer Research Center, Kashirskoye shosse, 24, Moscow, Russia, 115478

**e-mail:** ira-bio@yandex.ru, irina.v.manina@gmail.com

**Abstract**

Various approaches to increase efficiency of antitumor therapy by a combination of vaccinotherapy, chemotherapy and surgical excision of primary tumor nodes, and also the comparative analyses of therapeutic and preventive application of antitumoral vaccines were carried out in melanoma experimental model. It was postulated that preventive vaccination is able to prevent tumor incidence by 70 %. The combination of vaccinotherapy and surgical treatment of melanoma increases antimetastatic activity of vaccination by 43 %. We conclude the combined therapy would lead to more effective antitumor response.

INTRODACTION

Presently in clinical practice tumor immunotherapy methods are actively developed and implemented, in particular vaccinotherapy [1,5,7]. Applying inactivated tumor cells to patients can be compared to applying tumor antigens because it is believed that after injection tumor cells will be absorbed by antigen-presenting cells (APC), processed and presented as antigen peptides associated with MHC molecules. Presently investigations are performed worldwide on using antitumor vaccines produced from tumor cells [1,7]. For improving the immune response, tumor cells are transfected by the gene encoding a cytokine. GM-CSF gene is the most active [1,3,7].

The key factor stipulating that GM-CSF-producing vaccines are more immunogenic, is the ability of this cytokine to induce differentiation of earlier predecessor cells in to the most effective "professional" APC – dendrite cells. Under the influence of GM-CSF there are increased levels of IL-1 and TNF production by macrophages and thus local inflammation reaction develops and macrophages are activated, granulocytes and NK, NKT cells are additionally attracted to the injection area [8]. The results of clinical applications of GM-CSF-producing vaccines produced from transfected autologic tumor cells of patients proved that therapy is effective for treatment of human cancer [12]. Positive clinical effects were noted in 25-30% patients with melanoma at stages 3-4, that were irresponsive to other chemical therapies employed previously. At the same time antitumor vaccination alone proved to be not as affective for cancer therapy [7,9,10].

In order to increase the effectiveness of vaccinotherapy different immunotherapy modes and methods are employed in appropriate experimental in vivo models [6,10,15].

The application of present day combined chemical therapy allowed to increase treatment efficacy, but survival rates remained unchanged [11]. Introducing immunotherapy methods in the combined therapy contributes to increasing the length of remissions without supporting therapy.

Such drugs as cyclophosphamide, doxorubicin can significantly improve antitumor immunization effect [9,14]. It is based firstly on the ability of a number a chemotherapy drugs to suppress T-regulatory cells, the withdrawal of which increases response to many antigens including those of tumor origin.

Immunotherapy by GM-CSF-producing vaccine in combination with cyclophosphamide and doxorubicin has been employed clinically since 2003 [10,13].

Stable remission after surgical treatment of initial tumor node at early stages can be achieved in 80-90% patients. On the other hand, standard surgical removal of local invasive skin melanoma does not guarantee a long disease-free period without relapse (the absence of local exacerbation and distant metastasis).

For late stages of the disease, usually, it is necessary to combine different kinds of therapy, i.e. immunotherapy, chemical therapy and surgical treatment [2].

The aim of this research is to study various methods of increasing tumor therapy effectiveness by means of combining vaccinotherapy, chemical therapy and surgical removal of initial tumor nodes using the experimental melanoma models as well as to comparatively analyze therapeutic and prophylactic usage of antitumor vaccination.

MATERIALS AND METHODS.

The experiments were carried out on C57Bl6 male mice, 22-25 gram in weight; mice were obtained from Animal Center (Russian Cancer Research Center, Moscow, Russia). Animals were treated according to the ethical guidelines of the Animal Centre, Russian Cancer Research Centre, Moscow, Russia. These mice were subcutaneously inoculated with melanoma cells.

B16F10 mice melanoma cells were used, which are similar in their immune-phenotype to human melanoma model. Cultured tumors were measured twice a week, the tumor volume being calculated according to the next formula:

$V = D_{max} \times D_1 \times (D_{max}/2)$, where

$D_{max}$ is maximum tumor diameter,

$D_1$ is diameter perpendicular to the maximum one.

The cells express a number of tumor specific antigens and particularly express low levels or none of the MHC. For the vaccinotherapy BG melanoma B16F10 GM-CSF-producing clones were used. They were exposed to γ-rays by Agat-R using $^{60}$Co at the experimental therapy clinic of the N.N. Blokhin Russian Cancer Research Center. The irradiation dose was 100 gray.

Antitumor effect was evaluated based on by tumour growth retardation and mice survival dynamics. Antimetastatic effects were evaluated by weight and metastatic colony number in lungs according to standard techniques. Surgical operations were performed using gexanal anesthesia (100 mg/kg one time intra abdominal).

RESALTS AND DISCUSSIONS.

With employment of prophylactic vaccination as a monotherapeutic approach, with the following transplantation of B16F10 cells in to mice, an increase in survivability was shown up to 50,3±3,0 days (p<0,05) in comparison with the control group (29,7±4,0), 70% of mice in this group did not show significant increase in the development of tumors. The vaccine was injected in to animals seven days before the tumor transplantation.

In order to induce tumors mice were transplanted subcutaneously with $10^5$ B16F10 cells. During the vaccination every mouse was injected subcutaneously with $1 \times 10^6$ of antitumor whole cell GM-CSF-producing vaccine.

The dynamics of tumor growth in a typical experiment is shown in table 1.

**Table 1. Frequency of tumor growth development against the total number of mice in the group.**

| Time (days) | Control group | Prophylactic vaccination group |
|---|---|---|
| 7 day | 3(7) | 2(7) |
| 9 day | 6(7) | 2(7) |
| 11 day | 7(7) | 6(7) |
| 14 day | 7(7) | 4(7) |
| 18 day | 7(7) | 4(7) |
| 22 day | 6(6) | 5(7) |
| 25 day | 6(6) | 3(7) |
| 29 day | 5 (5) | 2(7) |
| 32 day | 3(3) | 2(6) |
| 37 day | - | 2(6) |
| 45 day | - | 1(5) |
| 52 day | - | 1(5) |

In the course of experiments appearance of infiltrates was noted at the place of injection, which disappeared by the end of the experiments. This process reflects the induction of the immune reaction at the place of injection, which can be described as delayed-type hypersensitivity reaction, and degree of its expression [4].

For treatment of developed tumors, vaccinotherapy as a monotherapy did not have any positive effect against the tumor growth or antimetastatic effect.

Interestingly, minor stimulation of tumor growth was noted. At the same time, we cannot rule out that in this case we deal with a "false" stimulation of tumor growth, associated with infiltration of the tumors with immune cells and stroma cells proliferating in tumor growth zone.

This statement is supported by the data provided in table 2, which shows the same life-span for the control group and the group that has received vaccination.

**Table 2. Therapeutic vaccinotherapy influence on subcutaneous melanoma B16F10 growth.**

| Influence | Overage life-time (days) | Life-time increasing (%) | Overage tumor mass (gr) | Tumor growth retarding (%) | Overage lungs mass (mg) | Metastatic process retarding (%) |
|---|---|---|---|---|---|---|
| Control group | 23,6±3,0 | - | 6,1±0,8 | - | 209,4±20,5 | - |
| Vaccination group | 24,1±3,5 | 2 | 11 ±1,9 | (+) 80 | 194,3±35,3 | 7 (2,0÷12,0) |

During the combined application of cytostatics, such as Cyclophosphamide, Cisplatinum, Doxorubicin, Lisomustine (in high and low therapeutic doses) and antitumor cell-whole GM-CSF-producing vaccine, no improvement in therapeutic effect against initial tumor node was recorded. However, increasing life-span of mice with melanoma B16F10 was noticed. The most effective type of combined therapy was using Cyclophosphamide in dose 30 mg/kg once a day during 5 days before and 5 days after vaccination. The dynamics of tumor growth suppression that developed in a typical experiment using Cyclophosphamide in dose 30 mg/kg is shown in fig. 1.

Application of complex therapy of vaccination and CPh was shown to inhibit tumor growth by more than 90%. By 25 day of the experiment (by the time of death of the control untreated mice) tumor growth suppression was 73%. Control animals have lived for 24,9±1,2 days, but complex treated animals have lived 37,0±5,1 days (p<0,05). Life-span increase was 49%, 95% CI (39,2÷58,8). Besides cytotoxic effects on tumor cells CPh has immune correcting effect. It is expressed by significant improvement of antitumor effects of the vaccinotherapy.

**Fig. 1. Tumor growth retarding using combine therapy: vaccinotherapy and Cyclophosphamide (CPh) for a long time.**

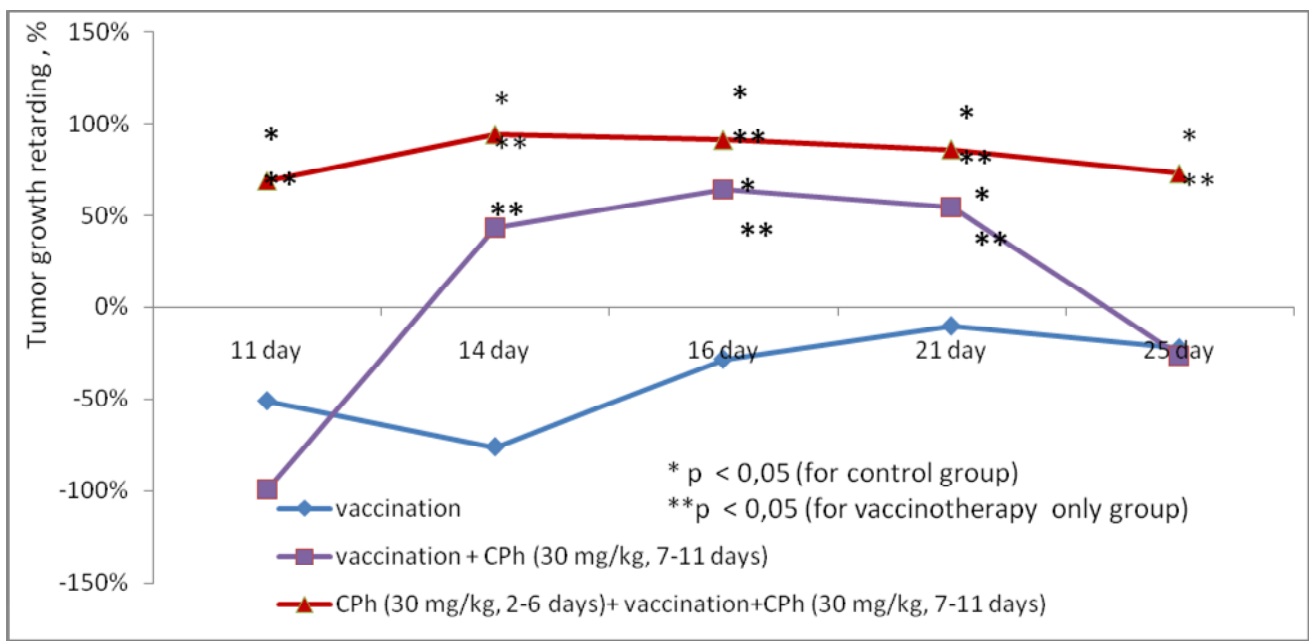

The vaccinotherapy also inhibited metastatic process of melanoma B16F10 by 29-43% using surgical excision of the initial tumor node (for various schedules of vaccination).

The most effective was schedule of vaccination before and after the surgical excision of the initial tumor node.

For biotherapy every mouse was injected with $1 \times 10^6$ antitumor cell-whole GM-CSF-producing vaccine cells subcutaneously.

For treatment of different animal groups, vaccination was performed prior (5 days before surgical excision of the initial tumor node), after (5 days after surgical excision of the initial tumor node), and prior/after the surgical excision of the initial tumor node (the same terms). Tumor metastization intensity was lower in vaccinated mice(down to 20-30%), than in the control group. The comparative analysis of vaccine antimetastatic activity using different surgical excisions of the initial tumor node schedules is presented in fig. 2.

**Fig. 2. The antitumor cell-whole GM-CSF-producing vaccine antimetastatic activity using different surgeon excision of initial tumor node regimens.**

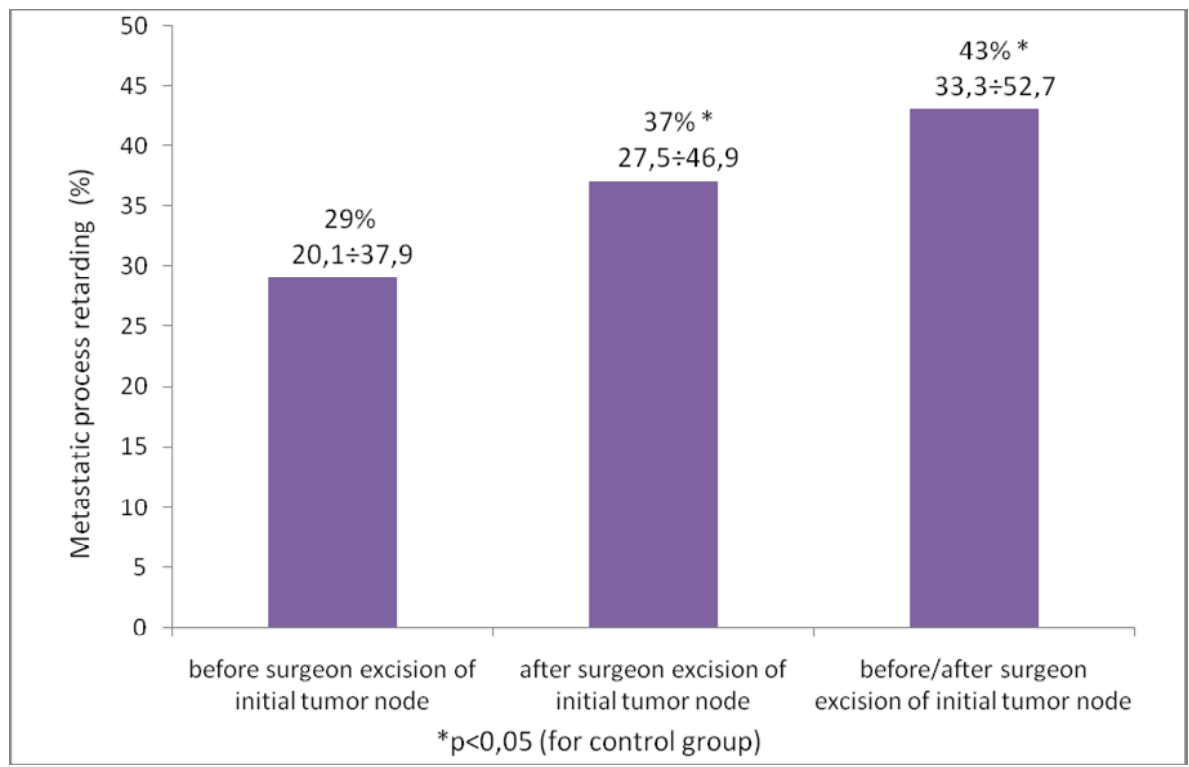

CONCLUSION.

1. It is useful and reasonable to combine various kinds of antitumor therapy for better therapeutic effets on tumor grouth.
2. It is shown that it is possible to use too kinds of antitumor agents: antitumor whole-cell GM-CSF-producing vaccine and cytostatics, such as Cyclophosphamide. As a result of the effectiveness of these therapeutic methods and their combines application it is possible expand the therapeutic possibilities on antitumor therapy.
3. Surgical treatment of the initial tumor node combined with vaccination twice produces an evident antimetastatic affect. Such results give the basis for optimization of vaccine therapy in clinical practice.

4. High prophylactic effectiveness of antitumor whole-cell GM-CSF-producing vaccine is an experimental ground for development of approaches to vaccination of melanoma patients.

*The researches was fulfilled under financial support of Moscow Government in the framework of scientific and technical program "Development and implementation in medical practice new methods and ways of diagnostics and treatment of oncological and other diseases".*

**References**

1. *Барышников А.Ю.* Принципы и практика вакцинотерапии рака // Бюллетень Сибирского отделения Российской академии медицинских наук. – 2004. – Т. 2. – С. 59-63.
2. *Барышников А.Ю., Демидов Л.В., Кадагидзе З.Г. и др.* Современные проблемы биотерапии злокачественных опухолей. // Вестник Московского онкологического общества. – 2008. – Т. 1. – С. 6-10.
3. *Бережной А.Е., Сапрыкина Н.С., Ларин С.С. и др.* Изучение противоопухолевой активности вакцин на основе генетически модифицированных опухолевых клеток, секретирующих ГМ-КСФ. // Российский Биотерапевтический Журнал. – 2006. – Т. 5, № 4. – С. 47-53.
4. *Михайлова И.Н.,Иванов П.В.,Петренко Н.Н. и др.* Внутрикожная клеточная реакция на фоне вакцинотерапии меланомы кожи.// Российский Биотерапевтический Журнал. –2010. – №1. – С. 63-67.
5. *Моисеенко В.М., Балдуева И.А.* Проблемы иммунологии опухолевого роста и возможности вакцинотерапии. //Медицинский академический журнал. –2007. – Т. 7, № 4. – С. 17-35.
6. *Berinstein NL.* Strategies to enhance the therapeutic activity of cancer vaccines: using melanoma as a model. // Ann N Y Acad Sci. – 2009.- Sep.-1174.- P.107-117.
7. *Borrello I., Pardoll D.* GM-CSF-based cellular vaccines: a review of the clinical experience. // Cytokine Growth Factor Rev. -2002.- Vol. 13.- P. 185-193.
8. *Dranoff G.* GM-CSF-secreting melanoma vaccines. // Oncogene.- 2003.- Vol. 22.- P.3188-3192.
9. *Emens LA., Armstrong A, Emens D et al.* A phase I vaccine safety and chemotherapy dose-finding trial of an allogeneic GM-CSF-secreting breast cancer vaccine given in a specifically timed sequence with immunomodulatory doses of cyclophosphamide and doxorubicin. // Hum Gene Ther. -2004.- Vol. 15.- P.313-337.

10. *Hege KM., Joss K, Pardoll. D.* GM-CSF gene-modified cancer cell immunotherapies: of mice and men. // Int Rev Immunol.- 2006.- Vol. 25.- P.321-52.

11. *Kohlmeyer J., Cron M., Landsberg J. et al.* Complete regression of advanced primary and metastatic mouse melanomas following combination chemoimmunotherapy. .// Cancer Res.- 2009 .-1Aug.-69(15).- P. 6265-6274.

12. *Li B, Simmons A, Du T et al.* Allogeneic GM-CSF-secreting tumor cell immunotherapies generate potent anti-tumor responses comparable to autologous tumor cell immunotherapies. // Clin Immunol.- 2009.- Nov.-133(2).- P. 184-197.

13. *Machiels JP, Reilly RT, Emens LA et al.* Cyclophosphamide, doxorubicin, and paclitaxel enhance the antitumor immune response of granulocyte/macrophage-colony stimulating factor-secreting whole-cell vaccines in HER-2/neu tolerized mice. // Cancer Res.- 2001.- Vol. 61.- P.3689-3697.

14. *Treisman J, Garlie N.* Systemic therapy for cutaneous melanoma. // Clin Plast Surg.- 2010 .-Jan.-37(1).- P. 127-146.

15. *Willem W, Overwijk N, Restifo P.* B16 as a Mouse Model for Human Melanoma.// Curr Protoc Immunol.- 2009.- 19 Oct .–20.1.- P.1-33.